\documentclass[aps,pra,floatfix,onecolumn,tightenlines,superscriptaddress,notitlepage]{revtex4-1}

\usepackage{amsfonts,amsmath,amssymb}
\usepackage{color}
\usepackage{graphicx}
\usepackage{multirow}
\usepackage[english]{babel}
\makeatletter\AtBeginDocument{\let\@elt\relax}\makeatother

\usepackage[colorlinks=true,citecolor=blue,linkcolor=blue,urlcolor=blue,pdftitle={Quantum Technologies in Space}]{hyperref}

\usepackage{longtable}
\usepackage{array}
\usepackage{enumitem}

\usepackage{color}                    			     
\begin{document}

\title{Quantum Technologies in Space}´

\author{Rainer Kaltenbaek}
\affiliation{Faculty  of  Mathematics  and  Physics, University  of  Ljubljana, Ljubljana,Slovenia}
\affiliation{Institute  for  Quantum  Optics  and  Quantum  Information, Vienna, Austria}

\author{Antonio Acin}
\affiliation{ICFO-Institut de Ciencies Fotoniques, The Barcelona Institute of Science and Technology, Cadtelldefels (Barcelona), Spain}
\affiliation{ICREA-Institucio Catalana de Recerca i Estudis Avan\c{c}ats, Barcelona, Spain}

\author{Laszlo Bacsardi}
\affiliation{Department of Networked Systems and Services, Budapest University of Technology and Economics, Budapest, Hungary}

\author{Paolo Bianco}
\affiliation{Airbus Defence and Space Ltd, Portsmouth, UK}

\author{Philippe Bouyer}
\affiliation{LP2N, Laboratoire Photonique, Numérique et Nanosciences, Université Bordeaux-IOGS-CNRS: UMR5298, Talence, France}

\author{Eleni Diamanti}
\affiliation{LIP6, Sorbonne Université, CNRS, Paris, France}

\author{Christoph Marquardt}
\affiliation{Max Planck Institute for the Science of Light, Erlangen, Germany}

\author{Yasser Omar}
\affiliation{Instituto Superior Técnico, Universidade de Lisboa, Portugal}
\affiliation{Instituto de Telecomunicações, Lisbon, Portugal}
\affiliation{Y Quantum -- Why Quantum Technologies Ltd., Portugal}

\author{Valerio Pruneri}
\affiliation{ICFO-Institut de Ciencies Fotoniques, The Barcelona Institute of Science and Technology, Cadtelldefels (Barcelona), Spain}
\affiliation{ICREA-Institucio Catalana de Recerca i Estudis Avan\c{c}ats, Barcelona, Spain}

\author{Ernst Rasel}
\affiliation{Institute for Quantum Optics, Leibniz University Hannover, Hannover, Germany}

\author{Bernhard Sang}
\affiliation{OHB System AG, We\ss{}ling, Germany}

\author{Stephan Seidel}
\affiliation{Airbus Defence and Space GmbH, Taufkirchen, Germany}

\author{Hendrik Ulbricht}
\affiliation{Department  of  Physics  and  Astronomy, University  of  Southampton, Southampton, UK}

\author{Rupert Ursin}
\affiliation{Institute  for  Quantum  Optics  and  Quantum  Information, Vienna, Austria}

\author{Paolo Villoresi}
\affiliation{Department of Information and Engineering, University of Padua, Italy}
\affiliation{Padua Quantum Technologies Research Center, University of Padua, Italy}

\author{Mathias van den Bossche}
\affiliation{Thales Alenia Space, Toulouse, France}

\author{Wolf von Klitzing}
\affiliation{Institute of Electronic Structure and Laser, Foundation for Research and Technology - Hellas, Heraklion, Greece}

\author{Hugo Zbinden}
\affiliation{University of Geneva, Geneva, Switzerland}

\author{Mauro Paternostro}
\affiliation{Centre  for  Theoretical  Atomic, Molecular  and  Optical  Physics, Queen’s  University Belfast, Belfast, UK}
\email[Correspondence e-mail address: ]{m.paternostro@qub.ac.uk}

\author{Angelo Bassi}
\affiliation{Department  of  Physics, University  of  Trieste, Trieste, Italy}
\affiliation{Istituto  Nazionale  di  Fisica  Nucleare, Trieste  Section, Trieste, Italy} 
\email[Correspondence e-mail address: ]{abassi@units.it}

\begin{abstract}
Recently, the European Commission supported by many European countries has
announced large investments towards the commercialization of quantum technology
(QT) to address and mitigate some of the biggest challenges facing today’s digital era --
e.g. secure communication and computing power. For more than two decades the QT
community has been working on the development of QTs, which promise landmark
breakthroughs leading to commercialization in various areas. The ambitious goals of
the QT community and expectations of EU authorities cannot be met solely by
individual initiatives of single countries, and therefore, require a combined European
effort of large and unprecedented dimensions comparable only to the Galileo or
Copernicus programs. Strong international competition calls for a coordinated European effort
towards the development of QT in and for space, including research and
development of technology in the areas of communication and sensing. Here, we aim at
summarizing the state of the art in the development of quantum technologies which
have an impact in the field of space applications. Our goal is to outline a complete
framework for the design, development, implementation, and exploitation of quantum
technology in space. 
\end{abstract}


\maketitle

\section{Introduction}\label{introduction}
In an increasingly connected and digitalised society the reliance on satellites, and more broadly on space, is crucial for the well-functioning of our daily lives. The limitations and the challenges that the current generation of technologies face have called authorities to invest more in promising emerging technologies, such as quantum technologies (QT).

Through the Quantum Technology Flagship (QT Flagship), the European Commission has identified four application areas that should be underpinned by research in the basic science enabling quantum technologies. All these areas have immediate applications also to Space science and industry.

Following the QT Flagship, the long-term vision that should be pursued is to integrate the terrestrial quantum web with a space one, where quantum computers, simulators and sensors are interconnected via quantum communication networks.

This white paper aims at summarizing the state of the art in the development of quantum technologies and their impact on the field of space applications, and to also delineate a roadmap for the consideration of major actors in this area, i.e. from the European Commission -- as responsible in the definition of the EU space strategy -- to ESA, national space agencies and industries.

\section{Secure Quantum Communication -- QC}
Quantum Communication (QC) uses the transfer of quantum information between distant terminals (see e.g. Refs.~\cite{Krenn2016Quantum,Muralidharan2016Optimal}. One of its possible uses is quantum key distribution (QKD)\cite{Diamanti2016Practical,Laudenbach2018Continuous-Variable,Lo2014Secure,Vallone2015Experimental}, which counters the threat posed by quantum computers on the widely used asymmetric encryption and will lead to long-term secure communication. The quantum secure systems developed so far in Europe provide secure communication on ground. The extension to space will be the necessary complement to build networks of distant nodes. Essential quantum secure solutions have already been envisaged and partially developed but much more is needed in Europe for more advanced, widespread applications, including daylight QKD\cite{Agnesi2020Simple,Avesani2019QCoSOne}, and research programs\cite{Agnesi2019All-fiber}. Beyond secure communication, general-purpose Quantum Information Networks will allow connecting quantum devices such as processors and sensors for tremendous increases of performance, and will open the door to the widest range of applications yet to be imagined\cite{Kozlowski2019Towards}.

\subsection{Quantum adds to security}
Quantum Communication changes the paradigm with respect to current secure communication standards, making the transfer of information secure for long-term requirements and protecting against potential attacks, including those expected from quantum computers. 

QKD consists of the establishment of secret keys in a way that is ``information theoretically secure''. This means that the security of QKD protocols can be proven mathematically, based on direct information-theoretic arguments\cite{Renner2005Information-theoretic}, eliminating the need for computational assumptions. There are two types of protocols for this, namely ``prepare and measure'' (PM-QKD) and ``entanglement-based'' (ENT-QKD). 

QKD systems reached commercial-maturity level on terrestrial fibre links, and are based on PM-QKD, among which Decoy-State Bennett-Brassard (DS-BB84) is the most common\cite{Bennett1984Quantum,Lo2005Decoy}. Key exchange exceeding a million secure bits per second was demonstrated on fibre links. PM-QKD is currently under development for space channels, with one project using a satellite in low-Earth orbit\cite{Liao2017Satellite-to-ground}. Such links, whose nodes -- dubbed trusted -- will allow the storage of keys for some time, will enable the realization of networks for key exchange of any size. 

ENT-QKD is also possible with the use of a photon-pair source in between the two terminals where the key is generated. This kind of protocol is based on quantum entanglement distribution. It allows communication of arbitrary quantum states, and thus provides a more generic means of communication called quantum information network (or sometimes ``quantum internet''), which can offer security and other applications beyond PM-QKD. As far as space is concerned, it allows establishing secret keys on the ground, for instance with the E91\cite{Ekert1991Quantum} or the BBM92\cite{Bennett1992Quantum} protocols, with the satellite just distributing the entangled pairs to ground terminals. This was recently demonstrated using the Micius satellite\cite{Yin2020Entanglement-based}. Yet the implementation of entanglement-based protocols remains much more challenging, all the more when making use of long fibres or of high-altitude space-based optical links. For that reason, entanglement-based protocols might be a solution for the longer term eventually yielding a broader set of enabled applications.

Space-based QKD will be a crucial element to bootstrap the quantum information network at the planetary scale and beyond. Both types of QKD protocols shall be used, based on the needs of specific applications being considered, with a natural progression from PM-QKD -- which is already well developed on the ground -- to ENT-QKD. Space-based quantum communication is expected to unlock the implementation of quantum-communication applications beyond quantum cryptography, for instance in the distribution of time, metrology, and distributed quantum computation, in addition to fundamental investigations.

\subsection{Security needs certification and accreditation}

Although security can be quantified on the basis of QKD measurements, this is not sufficient for an operational system with mission-critical uses. In order to evaluate the security of a system, its use-case, boundary conditions and requirements need to be defined. The design and development of such a system shall then comply with rules derived from these requirements, and its compliance with them continuously assessed along the process.

In the case of quantum cryptography, an interdisciplinary approach is required between quantum physicists, cryptography/security experts, and engineers. Standards need to be defined and certification procedures developed by the relevant agencies on both national and international levels. Moreover, quantum and classical security concepts need to be combined and assessed in a single framework. These topics are currently being addressed by working groups within major standardization institutes, namely the European Telecommunications Standards Institute (ETSI), the International Standards Organization (ISO), and the International Telecommunication Union (ITU), for QKD as well as for longer-term quantum information networks. 

\subsection{Secure communication needs a system}

The design of a secure communication infrastructure requires a complete system approach including space and terrestrial components. The system design has to be adapted to different use-cases. On ground, fibre-based solutions will offer short-range secure communication. Satellite-based quantum communication will provide a means for reaching global distances and secure space assets. The technology of terrestrial components (ground stations and management centers) and space components (satellite or satellite networks) need to be adapted to different use-cases.

The deployment of keys between receivers on ground to be used in pan-European as well as national secure communications will be the most immediate results of such Space QKD system. The development of compact receivers, suitable for the roof of a normal building, would allow the demonstration of QKD to all Member States regardless of their geographical location, much before the development of a continental repeater network based on fibres.

\subsection{A roadmap for space-based secure quantum communication}

Quantum communication should be implemented on a practical and reliable platform to provide world-wide secure communication for European assets. This is needed as soon as possible, as is the European-scale demonstration and the involvement of all Member States in the QKD network. As the time for implementation and in-orbit operation of the usual satellite-based systems goes well beyond a decade, and the threat of a future quantum computer has to be considered in the same time-frame (even sooner in case of retroactive decryption), the urgency of action is striking.

Space-based QKD can provide access to such networks sooner and easier in the case of Member States or their territories that are geographically separated from others by long distances or by sea. The pursuit of such goals defines a clear roadmap that includes the following key pillars, to be developed in parallel:

Operational systems. SAGA (Security And cryptoGrAphic mission)\cite{Lewis2019secure} was originally an ESA internal study for ENT-QKD. In this context, we use SAGA as reference to the space component (encompassing both ground and space segments) of the future Quantum Communication Infrastructures (QCIs), comprising different types of implementation:
·	SAGA1: ``Security \& cryptography'': The aim is to develop satellite-based quantum cryptography for world-wide secure key-distribution using PM-QKD protocols and networks with trusted nodes. The system should include key management, secure symmetric encryption on classical channels, and options depending on user needs. Terrestrial terminals (ground receivers for the keys and interfaces to the fibre links) have to be developed as economically affordable, easily deployable and that can connect to ground-based fibre networks. (Demonstration mission launch: 2023/2024; Full Service starting: 2028/2029.)
·	SAGA2: ``Further applications'': entanglement-based and more. The aim is to develop entanglement-based quantum communication using satellites. Depending on use-cases the satellite may use different types of orbits. In addition, quantum repeaters may concretely be implemented in space systems to enable further applications on a larger scale. As a vision, a full quantum information network can be implemented world-wide. (Demonstration mission launch: 2028; Entanglement Distribution Service: 2035.)

Standards/certification development. Standardization and certification efforts have to start in parallel to the development of the operational systems.  First, the development of operational systems will serve as an input to the development of standards and certification, and later the standards and certification requirements will steer the development of the operational systems. (First draft standards: 2021; First certification: 2022; Advanced certification and standards developed: 2024; Certification for service: 2028.)

\begin{figure}[b!]
\centering
\includegraphics[width=0.8\linewidth]{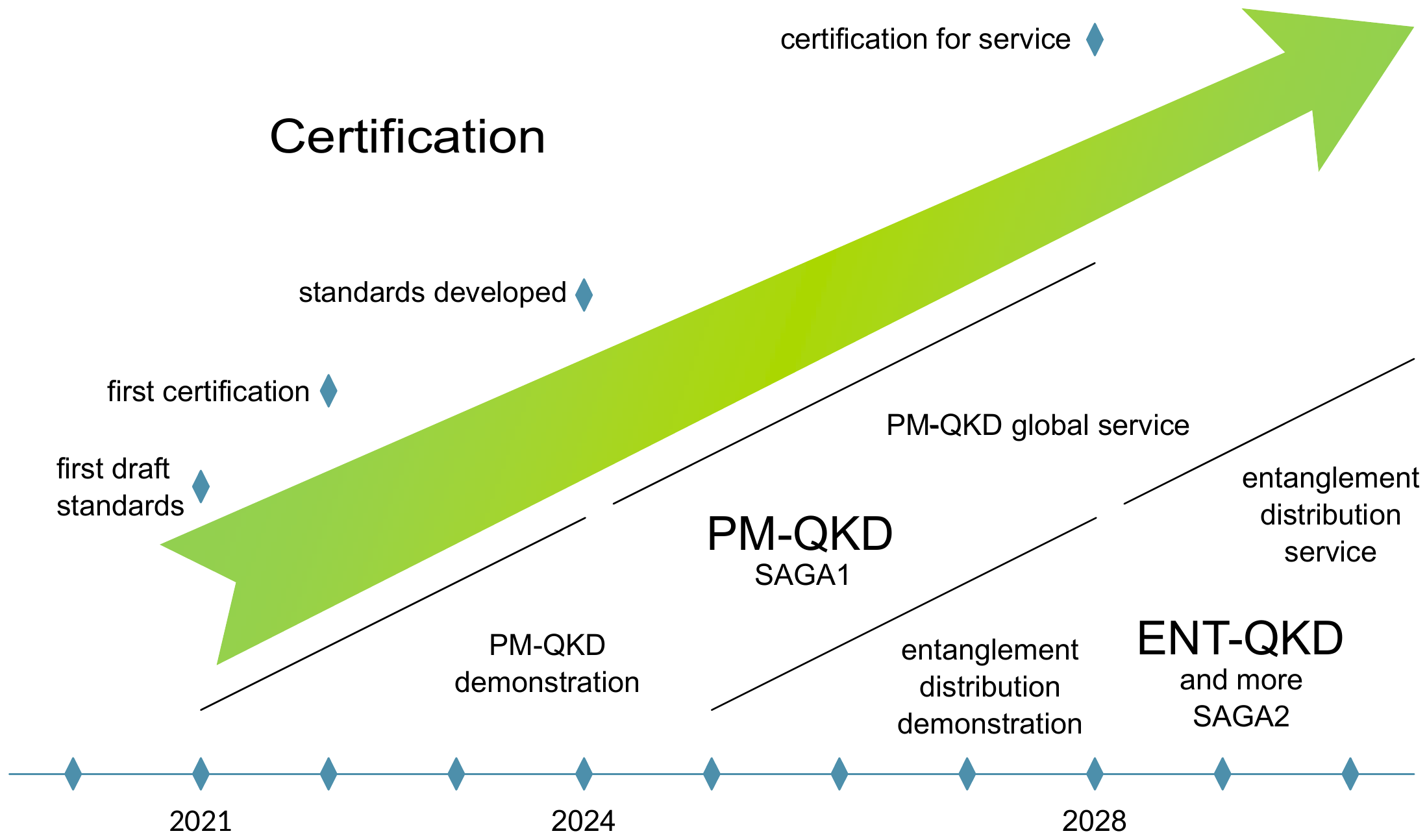}
\caption{The envisioned timeline for the proposed roadmap.}
\label{fig::roadmap}
\end{figure}

This vision is encompassed in Figure \ref{fig::roadmap}.

\subsection{R\&D on security, quantum concepts, and system concepts}

In addition to the concrete operational developments, research and development is needed in interdisciplinary topics. These include the development of high-rate QKD payloads, advanced pointing systems, low-loss space-to-ground links, also based on adaptive optical systems, and inter-satellite links.

Moreover, we aim at the exploitation of quantum functionalities where a provable advantage can be shown and that require current, near-term, or longer-term quantum technology. The most notable ones are:

\begin{itemize}
\item Quantum random number generation, which is achievable with current or near-term technology.
\item Quantum communication complexity protocols (with the most prominent ones being quantum fingerprinting and hidden matching/sampling matching).
\item Quantum cryptographic primitives like position verification, oblivious transfer, coin flipping, anonymous message transmission, leader election, secret sharing.
\item Delegated and distributed quantum computing.
\end{itemize}

In our opinion, the EU effort in this regard is essential to complement that of the Space Agencies.

The exploration of QKD protocols using different photonic degrees-of-freedom (for example, frequency, time bins, polarization, orbital angular momentum) will require advanced security analysis and the development of frameworks enabling new, affordable and practical services, and the highest security applications as well as a variety of new applications involving time, position, navigation, and much more.

\subsection{A vision for financial support}

To develop the program described here, we envisage a joint effort of the European Space Agency and the European Commission. To establish the space part of the European institutional QKD infrastructure, we recommend that ESA shall act as project management for the European Commission for technological and program implementation, as is the case for Europe's global navigation system, Galileo. In addition, we recommend that development of protocols and R\&D continue being supported by the Commission's Directorate-General for Defence Industry and Space (DEFIS)  and the Directorate-General for Communications Networks, Content and Technology (DG Connect) directly.

\section{Time and Frequency Transfer -- TFT}

Time standards and frequency transfer (TFT) are fundamental for many modern day applications with high societal value. Fundamental TFT techniques are well established today enabling services in the fields of communication, metrology and the Global Navigation Satellite System (GNSS). With the availability of optical atomic clocks and optical frequency transfer, QT enables the TFT performance to be boosted by several orders of magnitude\cite{Jozsa2000Quantum,Giovannetti2001Quantum-enhanced,Ilo-Okeke2018Remote}. This increased performance permits keeping pace with the steadily increasing needs in communications (timekeeping) and GNSS (geolocation), and enables new applications (geodesy, gravitational wave observation, synthetic aperture optical astronomy). A space component is relevant to the enhanced applications by enabling long range transfer, enhanced security and global availability.

\subsection{Time standards and frequency transfer-based applications}

Time standards and frequency transfer have a rich set of applications relevant to engineering, science and society. In high-speed communications, timekeeping makes use of atomic clocks synchronized by GNSS based time dissemination services to perform time stamping and information routing. These capabilities are fundamental to the high performance and availability of the internet and all its associated services. GNSS based services rely on high performance TFT capabilities. They have spawned many services and applications leading to multi-billion Euro turnover per year, and are key to defense and security today as well as enabling for future applications such as autonomous driving\cite{2017Springer}. 

Precision time standards are at the root of the aforementioned applications and modern day metrology. Precision time metrology is at the heart of the international standard definition and an enabling factor in a trade-based economy. Recently, in radio astronomy real-time imaging of a black hole has been demonstrated using synthetic aperture imaging for radio frequency signals\cite{EventHorizonCollaboration2019FirstA,EventHorizonCollaboration2019FirstB}. This was enabled by TFT technologies.

Fundamentally TFT is based on two technological elements: precision time standards (clocks) and the ability to transfer frequency (more precisely the phase of the clock) over a large distance with high precision. State-of-the-art commercial atomic clocks operate at radio frequencies (several GHz) and achieve accuracies down to $10^{-15}$. Current time standards are defined on improved clocks of the same type with accuracies of about $10^{-16}$\cite{Cartlidge2018Better}. The limitation of these clocks is given by the fundamental frequency of operation (GHz) and the ability to measure these. Phase (and frequency) transfers are standard methods in high frequency RF communications and have been demonstrated on satellite links with a performance of $10^{-15}$\cite{Riehle2017Optical} -- with the same fundamental limitations imposed by the frequency used for the transfer. These technologies are about to be demonstrated at increased accuracy in space in ESA’s International Space Station-based Atomic Clock Ensemble in Space (ACES) project, which contains both frequency standards and transfer capabilities to ground\cite{Savalle2019Gravitational}. ACES is currently scheduled to launch in mid-2021\cite{Sun2020Test}.

\subsection{Quantum enhanced performance, quantum enabled applications}

The fundamental improvement enabled by QT is given by an increase in fundamental frequency from the radio frequency domain into the optical domain by a factor of $10^5$\cite{Ludlow2015Optical}. This step is possible based on the evolution of laser technology (stable optical oscillators and frequency combs) and optical clock development (quantum state control). Optical clocks have been demonstrated in the laboratory environment to have a fractional uncertainty of less than $10^{-18}$\cite{Brewer2019} with further improvements expected (optical lattice clock and single ion clock). Frequency transfers have been demonstrated on a ground-based optical fibre link with a length of $920\,$km at an accuracy of better than $4\times 10^{-19}$\cite{Droste2012Optical}. 

ESA has studied the implementation of a sequel mission to ACES based on an optical atomic clock and optical links concept, Space Optical Clock on the International Space Station (ISS, I-SOC) with a possible implementation in the early 2020s\cite{Origlia2016Development}. These experiments clearly show the superior performance of the QT-based optical TFT which have improved the performance at this early stage of development already by three orders of magnitude (two in stability, one in precision)\cite{Origlia2018Towards}.

Making use of the technological advancement will bring improvements to the quality of existing applications or allowing building more efficient system architectures. Time dissemination services and metrology will have a direct improvement of three orders of magnitude. High precision clocks in space will provide a secure (i.e. hard to jam) and independent time base for global time keeping. Combined with space-space and space-ground optical links they will allow global TFT. Such a space-based time standard and time distribution system will provide an efficient globally available infrastructure (compared to fibre networks which are only available in densely populated areas) for all of the above applications. For Global Navigation Satellite (GNS) Systems, the system architecture is suggested to be improved with a more efficient implementation than is currently possible and a higher accuracy for specific use-cases.

With the accuracy of optical clocks and the capability to compare these at large distances, new applications now become feasible. As the relative gravitational red shift is $10^{-18}$ per cm of geopotential height, clocks and optical transfers can be used to measure the geopotential difference of two locations\cite{McGrew2018Atomic}. If one location is, for example, in orbit, the absolute geopotential on Earth’s surface can be characterized (geodesy application). Similar to this application gravitational waves can be detected by comparing two optical clocks on two distant satellites\cite{ElNeaj2020AEDGE,Kolkowitz2016Gravitational,Tino2019SAGE}. This concept augments the ESA/NASA Laser Interferometer Space Antenna (LISA) system in the fact that it has very high sensitivity to gravitational waves at low frequency, i.e., it is complementary to LISA. Finally, analogous to the radio-frequency synthetic-aperture observation concept, an optical synthetic aperture telescope can be envisaged by the fact that a set of fully synchronized optical clocks enables a phase measurement of the impinging light wave at several locations with large separation. This allows ``synthesizing by analysis'' a telescope with an aperture size comparable to the separation of the locations (potentially thousands of kilometres) leading to the capability of direct observation of extrasolar planets.

\subsection{Roadmap for QT enhanced TFT}

The following roadmap for the maturation of QT enhanced TFT technologies is suggested based on the maturity and complexity of the individual elements.
\subsubsection*{Short-term goals (5 years)}
As the technology elements for frequency transfer are readily available for fibre and free-space links and several applications are possible using frequency transfer without space-based optical clocks (geo-potential comparison, clock comparison) it is suggested to establish an infrastructure capable of frequency comparisons at $10^{-18}$ level accuracy. The TRL for this technology is currently evaluated at level 4--5.
In preparation of an optical clock in space the selection of clock concepts and technology shall mature.
\subsubsection*{Medium-term goals (10 years)}
Building on the results of optical link demonstration an improvement of one order of magnitude in frequency transfer shall be achieved ($10^{-19}$).
An in-orbit technology demonstration of the selected optical clock concept with a goal accuracy of $10^{-18}$ shall be realized.
\subsubsection*{Long-term goals ($>$ 10 years)}
With the QT enabled TFT technologies now at sufficient accuracy for many applications and with a TRL greater than 6 several missions can be realized:
\begin{itemize}
\item Universal time dissemination
\item Geodesy service
\item New GNSS architectures
\item Gravitational wave detection at low frequency
\item Fundamental physics experiments
\item Optical synthetic aperture telescopes
\end{itemize}

\section{Earth Sensing and Observation -- EO}

Gravity field mapping from space provides crucial information for the understanding of climate change, hydro- and biosphere evolution, and tectonics and earthquake prediction. The recent advent of macroscopic quantum matter such as Bose-Einstein condensates and the associated Nobel-prize winning protocols, have led to inertial quantum sensors based on atom interferometry. Quantum gravity sensors use coherent quantum matter waves as test masses, which lead to far more sensitive and precise instruments. Space-based quantum sensors will enable better monitoring of the Earth’s resources and improve the predictions of Earth-quakes and the adverse effects of climate change, like draughts and floods. 

\subsection{The need for Space-Based Quantum Sensing}

Given the extreme effects of global warming that humankind is facing, Earth observation is perhaps the most important scientific endeavour of our times. Already today, the study of global mass transport phenomena via satellite gravimetry provides important insights for the evolution of our planet and climate change, by improving our understanding of the distribution of water and its changes.  Recently, the NASA gravity mission, Grace, found that the temperature of the water in the deep ocean rifts has not changed in recent decades\cite{Wouters2014GRACE}. Gravimetric satellite data are now used as an early warning system for floods as well as droughts both in the agriculturally important Midwest of the US\cite{Thomas2016comparison} and in the rainforest in Congo\cite{Crowley2006Land}. ESA’s gravity mission (GOCE)\cite{Cesare2010measurement} produced detailed maps of deep-sea ocean currents, which are very important drivers of Earth’s climate. It revealed the remnants of lost continents hidden deep under the ice sheet of Antarctica, and it also detected a post-glacial rebound. The changes of Earth's gravity scarred by earthquakes gives valuable data for predicting future earthquakes. GOCE’s data are also being used to improve models of Earth's geology, indicating the potential locations of subsurface energy sources. Just like gravitational waves heralded a new age in astronomy, mapping the earth’s gravitational field is proving to be an immensely valuable tool in understanding Earth\cite{Loomis2020Replacing}.

ESA and NASA are working on future versions of their gravity missions, which will have much improved resolution and sensitivity.  However, it has become clear that the classical measurements cannot be pushed much further. Classical accelerometers used so far in gravity missions exhibit increased noise at low frequency and have large long-term drifts. This severely limits the ability of faithfully reconstructing the earth gravity field at low degrees and precisely modelling its temporal fluctuations. The NASA future geodesy mission, Mass Change (MC), is in preparation for launch by 2026\cite{Caltech2019Mass}. 

\subsection{Quantum mapping of the Earth’ mass dynamics}

Today, the study of global mass transport phenomena via satellite gravimetry provides important insights for the evolution of our planet and climate change\cite{Kornfeld2019GRACEfo}. Atom interferometry will play an instrumental role to improve satellite-based measurements for space-geodesy. They promise smaller low-frequency noise and long-term drifts, which will be beneficial for faithfully reconstructing the Earth’s gravity field at low degrees and even more for precisely modelling its temporal fluctuations. 

Atom-interferometric quantum sensors offer far superior long-term stability and higher sensitivity. For this reason, quantum sensors are considered by ESA as a potential instrument\cite{Mueller2020Cold}, or a demonstrator\cite{Bidel2020}. ESA’s future geodesy mission, Next Generation Gravity Mission\cite{Haagmans2020ESA}, classified as a Mission of Opportunity, will include laser ranging but consider quantum sensors as a candidate for the following mission if the technology is ready at that time.

\subsection{Europe’s pioneering of space quantum sensing}

Since 2000, missions exploiting inertial quantum sensors (IQS) have been proposed to ESA. In 2010, an ESA road map highlighted space-borne IQS for fundamental physics in space, such as gravitational wave detection. IQS were selected for studies by ESA for satellite gravimetry using cold-atom interferometry (CAI)\cite{Bresson2006Quantum}, the Lense-Thirring effect with the hyper-precision cold-atom interferometry in space proposal (HYPER)\cite{Jentsch2004HYPER}, as transportable devices in the space atom interferometer proposal (SAI) and, most recently, was among the three selected candidate missions for a quantum test of the equivalence principle (STE-QUEST: Space-Time Explorer and QUantum Equivalence principle Space Test)\cite{Hechenblaikner2014STEquest}. Concerns of technological immaturity prevented selection and motivated support for some critical payload sub-components. 

Since the first decade of this century, national agencies have supported space-borne quantum-sensor development. Important milestones were achieved by (ICE (Interf\'{e}rom\'{e}trie atomique \`{a} sources Coh\'{e}rentes pour l'Espace, France) on parabolic flight studies for dual-species interferometry\cite{Varoquaux2009How}, by QUANTUS (Quantum Gases in Weightlessness, Germany) and MAIUS (Matter-Wave Interferometry in Weightlessness, Germany) establishing interferometry with Bose-Einstein condensates in space based on drop-tower\cite{Zoest2010bec} and sounding-rocket experiments\cite{Becker2018Space}, and by the cold-atom laboratory (CAL, USA)\cite{Elliott2018NASA}, where US teams including German scientists explore physics with Bose-Einstein condensates in orbit in the Bose Einstein Condensates and Cold Atoms Lab (BECCAL)\cite{Frye2019bec}. Benefiting from the space activities, novel terrestrial sensors and commercial spin-offs were created and will continue to emerge as industry starts to engage in the development of cold atom payloads.

\subsection{Worldwide Earth observation with European quantum technologies}

Galileo exemplifies Europe’s ambition to establish its independence regarding key technologies for space. In the recent past, stepping-stones were placed in developing methods for space-borne high-precision gravity sensing. These sensors promise to improve Earth observation by their long-term stability and low drift. The grip of gravity is compromising the analysis of the performance of these sensors. Microgravity facilities provide only limited access to extend the free-fall and space-like environment for raising the TRL of key components and method development; only a pathfinder can sound out the potential performance in the relevant environment. The European competences have to be firmly bundled to master this challenge and to establish space-hardware for such a sensor. Mission design and exploitation plans need a close interdisciplinary cooperation between the quantum sensor community, high-tech industry, and geodesists. At this point, firm commitments between partners and stakeholders are required to jointly prepare such a mission.
 
\subsection{A roadmap for quantum sensors for Earth sensing and observation}
\begin{enumerate}[label=(\alph*)]
\item This requires a strong coordination in prototyping and performance tests. This could, for example, be performed in a joint research coordination as has been proposed by ESA technical officers for some time in the ``C-COOL'' laboratory initiative. Such a joint laboratory could harness European ground-based microgravity facilities and expertise and exploit European heritage and synergies. Parallel to this, the initiatives of ESA, CNES and DLR to elaborate the most efficient mission concept for satellite gravimetry with quantum sensors should be consolidated, especially with respect to improving ultimate performances (enhanced sensitivity, AOCS, size, mass and power reduction). All these efforts must be supported with the objective of constructing an Engineering Qualification Model, e.g., a space prototype that will also be used as the payload on a small pathfinder satellite to allow for a rapid launch (5 years).
\item A model of an IQS for space needs to be developed and engineered. An Earth-Venture-like mission should be designed as pathfinder for in-orbit validation. This will start the crucial process of transfer of know-how to industry, the development of space-qualified hardware, and of an elegant breadboard for prototyping and performance tests in microgravity (10 years).
\item The final goal is to perform a geodesy mission using one or multiple space-borne quantum sensors; exploit quantum sensors for other applications such as navigation, exploration and planetology (Moon, Mars) (> 10 years). 
\end{enumerate}

\section{Fundamental Physics -- FP}
Space is an exquisite environment for unique experimental tests of the fundamental laws of nature: general relativity (GR), quantum mechanics (QM), Cosmology (Dark Energy and Dark Matter). The extension of fundamental tests using quantum correlations along unprecedented distances provides a unique testbed for fundamental tests\cite{Joshi2017Space,Kaltenbaek2004PoC,Vedovato2017Extending}. Moreover, cutting-edge technology can achieve extremely low-noise, low-gravity conditions in space. These advantages of a space environment promise the opportunity to push the boundaries of our understanding of Nature, in particular, for testing the existence of macroscopic quantum states. 

\subsection{Advancement of QT in space}

The Chinese Micius satellite, US-lead CAL on the ISS and German MAIUS rocket exploring Bose-Einstein condensates showed impressive QT-based results. In 2010, ESA published with community support a road map for fundamental physics highlighting the role of quantum technologies\cite{ESAroadmap}. Prominent fundamental-physics mission proposals such as STE-QUEST\cite{Hechenblaikner2014STEquest} and Macroscopic Quantum Resonators (MAQRO) \cite{Kaltenbaek2012Macroscopic,Kaltenbaek2016Macroscopic} were investigated by ESA with no clear elected candidate. HYPER\cite{Jentsch2004HYPER} and, recently the Quantum Physics Platform (QPPF)\cite{QPPF2019}, were selected for a pilot study by the concurrent design facility (CDF). STE-QUEST was selected as a candidate for a medium-sized missions within Cosmic Vision in the context of an ESA call for proposals in 2010. No proposed or pre-selected mission exploiting QT was finally nominated due to major concerns about technology immaturity and feasibility.

In the QPPF study, a payload was designed and optimized, based on earlier work for MAQRO\cite{Kaltenbaek2016Macroscopic} and a related proposal for near-field high-mass interferometry on ground\cite{Bateman2014NearField,Delic2020Cooling} in order to address three science objectives: (1) testing quantum physics in a regime where alternative theoretical models predict noticeable deviations from quantum physics, (2) quantifying any such deviations, and (3) the tested parameter regime should have an overlap with experiments possible on ground. The QPPF study showed that such experiments should be possible by using appropriate cryogenic cooling and thermal shielding of the instrument. Given dedicated technology development, the QPPF study concluded it will be possible to perform MAQRO-type experiments in space with test masses of at least up to $\sim 5\times 10^9$  atomic mass units. However, the QPPF study identified three critical issues that first need to be addressed: (1) realizing a reliable method to load the test particles into an optical trap in vacuum, (2) achieving a sufficiently high vacuum (lower than $10^{-13}\,$mbar, ideally lower than $10^{-15}\,$mbar), (3) the originally proposed method for preparing high-mass superpositions could lead to decoherence due to light scattering. Efforts are on-going to address these critical issues. A prominent example of the progress being made is the first ground-state preparation of optically trapped sub-micron particles\cite{Delic2020Cooling}.

\subsection{FP benefits from QT in space}

Nearly 30 years ago, LISA\cite{Amaro-Seoane2017Laser}, a long-standing L-class ESA mission to detect gravitational waves, faced similar scepticism for technological and even fundamental reasons. Starting with a low TRL of the employed sensing method, a dedicated pathfinder mission was created to demonstrate the appropriateness and performance of the proposed measurement concept. The LISA Pathfinder mission successfully completed in 2017\cite{Armano2016SubFemto,Armano2017ChargeInduced}. In this sense, LISA could be seen as a successful role model for the design of high-risk experiments to host QT in space where Europe has demonstrated leadership. 

Clearly, space is the only environment to enable some of the ambitious scientific goals such as the generation of macroscopic quantum superposition and entangled states -- the ultimate test of quantum mechanics.

\subsection{QT developments -- three platforms: photons, atoms, optomechanics}

Available QTs for a fully-fletched FP mission include: classical and non-classical optical interferometry, the generation of non-local superpositions and entangled states, and the demonstration of robust-in-the-field quantum information protocols with photons. Atomic clocks and interferometers resembling the most precise meters and clocks available that build on quantum superposition states are becoming ever more robust and compact.

Optomechanical systems and matter-wave interferometers are pushing the boundaries of macroscopic quantum states in laboratory environments. 

The interplay of relativity with quantum phenomena shall be studied in the framework of the observation of the strength of quantum correlations using photon states of different types, entangled or not. Space allows for experimental schemes that are impossible on the ground, in terms of gravitational potential variations, length scales and relative velocities, which widens the domain in which we are testing our current understanding of nature.

\subsection{A collaborative effort}

Budgetary constraints for M/L-class missions require collaboration -- across the scientific communities, industry, agencies, on an international level -- and the joint decision to select only a few topical FP missions and on their execution on a single or multi-platform approach. Europe has to be at the forefront in QT in order to launch a mission with ESA alone or in cooperation with strong worldwide partners. 

\subsection{Drawing a roadmap for FP with QT in space}

For the first 5 years, we propose to define a roadmap for various selected scientific objectives and to develop key QT methods for the payload in a joint European effort. The advance of QTs to the required TRLs in micro-g environments remains the major technical challenge.

In the medium term (5-10 years), the FP community needs support to develop demonstrators for tests in microgravity environment and has to define pathfinders for validation of QT in orbit. An important objective is to explore platforms for multiple pathfinder activities. 

Clearly, the long-term goal ($>$ 10 years) is to exploit QT for the scientific objective to test quantum mechanical states in an ESA mission with/without worldwide partners.

\subsection{Proposed actions}

We recommend to implement a dedicated sustainable European Union program (5-10 years) to develop QT for FP in space in the coming years for the purpose of funding European collaborative projects, in order to perform proof-of-principle tests (5 years) and for increasing required TRLs (10 years) in order to be competitive in ESA’s calls. It is clear, that we need to exploit ESA’s mission heritage to define and develop together with ESA sustainable (5-10 year) programs for the transfer of technical innovations to space industries.

We further propose the establishment of joint, international laboratories for research teams to assemble core skills and knowledge to develop and test QTs for space towards the required TRLs and prepare the FP space missions.

\section{Conclusions}

Quantum technology will lead to a large range of new applications in space applications, in high precision measurements and in tests of the core foundations of physics. The next years will see the establishment of quantum networking infrastructure in Europe and around the world with the goal of achieving secure quantum communication and to lay the foundation for global quantum networks. In the long term, this will allow connecting quantum computer, quantum simulators and quantum sensors worldwide. At the same time, optical atomic clocks will allow measuring time and frequency with unprecedented precision, allowing even better global navigation, new tests of the foundations of physics, and novel approaches to gravitational wave observation. Atom interferometers will allow high-precision gravimetry allowing for high-accuracy Earth observation, but they will also provide a versatile new tool to study other celestial bodies like the Moon or Mars by using gravimetry. The possibility to achieve excellent zero-gravity conditions and long free fall times in space provides an excellent opportunity to use quantum physics for high-precision measurements, for example, for gravitational-wave observation or for testing the very foundations of quantum physics itself.

\section*{Acknowledgments}
RK, CM, WK, YO, HU, MP, and AB acknowledge support by the COST Action QTSpace (CA15220). AA and VP acknowledge funding from European Union’s Horizon 2020 research and innovation programme under the grant agreement No 820466 (CiViQ). AA acknowledges financial support from the ERC AdG CERQUTE, the AXA Chair in Quantum Information Science, the Government of Spain (FIS2020-TRANQI and Severo Ochoa CEX2019-000910-S), Fundaci\'{o} Cellex, Fundaci\'{o} Mir-Puig, Generalitat de Catalunya (CERCA, AGAUR SGR 1381). RK acknowledges support by the Austrian Research and Promotion Agency FFG (projects 854036, 865996). WK acknowledges funding from the European Union's Horizon 2020 research and innovation programme H2020-FETOPEN-2018-2019-2020-01 under grant agreement No 863127 ``nanoLace'' and the contribution of the AtomQT COST Action CA16221. ED acknowledges funding from the European Union’s Horizon 2020 Research and Innovation Programme under Grant Agreements No. 820466 (CiViQ) and 857156 (OpenQKD). ER's contribution to the presented work is supported by the CRC 1227 DQmat within the project B07, the EXC 2123 Quantum Frontiers within the research units B02 and B05, the QUEST-LFS, the German Space Agency (DLR) with funds provided by the Federal Ministry of Economic Affairs and Energy (BMWi) due to an enactment of the German Bundestag under Grant No. DLR 50WP1431 (QUANTUS-IV-MAIUS), 50WM1952 (QUANTUS-V-Fallturm), 50WP1700 (BECCAL), 50RK1957 (QGYRO), and the Verein Deutscher Ingenieure (VDI) with funds provided by the Federal Ministry of Education and Research (BMBF) under Grant No. VDI 13N14838 (TAIOL). ER acknowledges financial support from ``Nieders\"{a}chsisches Vorab'' through ``F\"{o}rderung von Wissenschaft und Technik in Forschung und Lehre'' for the initial funding of research in the new DLR-SI Institute and through the ``Quantum- and Nano-Metrology (QUANOMET)'' initiative within the project QT3. AB acknowledges financial support from the H2020 FET Project TEQ (Grant No. 766900), INFN, FQXi and the University of Trieste. MP thanks the H2020-FETOPEN-2018-2020 TEQ (grant nr. 766900), the DfE-SFI Investigator Programme (grant 15/IA/2864), the Royal Society Wolfson Research Fellowship (RSWF\\R3\\183013), the Leverhulme Trust Research Project Grant (grant nr. RGP-2018-266), the UK EPSRC (grant nr. EP/T028106/1). LB thanks the support of the J\'{a}nos Bolyai Research Scholarship of the Hungarian Academy of Sciences and the support of the Ministry of Innovation and Technology and the National Research, Development and Innovation Office within the Quantum Information National Laboratory of Hungary. HU acknowledges financial support from the EU H2020 FET project TEQ (Grant No. 766900), the Leverhulme Trust (Grant No. RPG-2016-046), and the UKRI Research England SPRINT project SIGMA. PV acknowledges support from the Ministero dell’Istruzione, dell’Universit\`{a} e della Ricerca under the initiative ``Departments of Excellence'' (Law 232/2016). YO thanks the support from Funda\c{c}\~ {a}o para a Ci\^{e}ncia e a Tecnologia (Portugal), namely through project UIDB/50008/2020 and from project QuantSat-PT. PB acknowledges support from CNES through the ICE technology development program, and the GRICE and CARIOCA mission studies. C-COOL is supported by ESA for the elaboration of a scientific and technical roadmap.

\section*{List of acronyms}\label{sec:listAcronyms}
%\addcontentsline{toc}{section}{\nameref{sec:listAcronyms}}

\begin{longtable}{r|p{13.5cm}}
\endfirsthead
\hline
\textbf{Acronym} & \textbf{Meaning}\\
\hline
\endhead
\hline\hline
\textbf{Acronym} & \textbf{Meaning}\\
\hline
ACES & Atomic Clock Ensemble in Space (ESA mission) \\
ARTES & Advanced Research in Telecommunications Systems (ESA Program)\\
BECCAL & Bose Einstein Condensates and Cold Atoms Lab\\
BB84 & The most utilized protocol for QKD\\
CAI & Col Atom Interferometry\\
CAL & Cold Atom Laboratory (NASA mission on the ISS)\\
CDF & Concurrent Design Facility\\
DEFIS & Directorate-General for Defence Industry and Space\\
DG Connect & Directorate-General for Communications Networks, Content and Technology\\
ENT-QKD & ``entanglement''-based QKD\\
ETSI & European Telecommunications Standards Institute\\
HYPER & hyper-precision cold-atom interferometry in space (ESA mission)\\
ICE & Interf\'{e}rom\'{e}trie atomique \`{a} sources Coh\'{e}rentes pour l'Espace (CNES mission)\\
I-SOC & Space Optic Clock on ISS, (candidate ESA mission)\\
IQS & Inertial Quantum Sensor\\
ISO & International Standards Organization\\
ISS & International Space Station\\
ITU & International Telecommunication Union\\
LISA & Laser Interferometer Space Antenna (ESA mission)\\
LISA Pathfinder & ESA/NASA mission\\
MAIUS & Matter-Wave Interferometry in Weightlessness (QUANTUS space experiment)\\
MAQRO & Macroscopic quantum resonators (candidate ESA mission)\\
PM-QKD & ``prepare \& measure'' QKD\\
QC & Quantum Communication\\
QCI & Quantum Communication Infrastructure\\
QPPF & Quantum Physics Platform (ESA CDF study)\\
QUANTUS & Quantum Gases in Weightlessness (research project)\\
QKD & Quantum Key Distribution, the most popular quantum cryptographic protocol\\
QT & Quantum Technology, Quantum Technologies\\
SAGA & Security And cryptoGrAphic mission (ESA ARTES precursor mission study)\\
STE-QUEST & Space-Time Explorer and QUantum Equivalence Principle Space Test (candidate ESA mission)\\
                                                    
\hline\hline
\end{longtable}

%\addcontentsline{toc}{section}{References}

%\bibliographystyle{abbrvnat}

%\bibliographystyle{apalike}
\bibliographystyle{apsrev4-1}

\bibliography{QTSpaceReferences}

\end{document}